# Stochastic Model Predictive Control of Air Conditioning System for Electric Vehicles: Sensitivity Study, Comparison and Improvement

Hongwen He, Hui Jia, Fengchun Sun, and Chao Sun*

*Abstract*—A stochastic model predictive controller (SMPC) of air conditioning (AC) system is proposed to improve the energy efficiency of electric vehicles (EV). A Markov-chain based velocity predictor is adopted to provide a sense of the future disturbances over the SMPC control horizon. The sensitivity of electrified AC plant to solar radiation, ambient temperature and relative air flow speed is quantificationally analyzed from an energy efficiency perspective. Three control approaches are compared in terms of the electricity consumption, cabin temperature, and comfort fluctuation, which are (i) the proposed SMPC method, (ii) a generally used bang-bang controller and (iii) dynamic programming (DP) as the benchmark. Real solar radiation and ambient temperature data are measured to validate the effectiveness of the SMPC. Comparison results illustrate that SMPC is able to improve the AC energy economy by 12% than rule-based controller. The cabin temperature variation is reduced by over 50.4%, resulting with a much better cabin comfort.

## I. INTRODUCTION

The task of air conditioning system in a car is to provide a thermally comfortable environment for the cabin, and is one of the most important auxiliary devices in modern vehicle. However, the AC system is also the most energy consuming unit besides the propulsion system [1]. Especially, for an electric vehicle (EV), the overall electricity consumption by AC system can be over 12%~30% of the total during a typical driving mission, and even higher [2]. Developing new methods to improve the operation efficiency of AC systems is very beneficial in extending EV's driving range [3-5].

In the literature, reducing the condensing temperature of the refrigerant, adding heat pumps or expansion valves, and enhancing the heat preservation abilities of the vehicle have been studied to improve AC system efficiency in automotive [6, 7]. Ünal et al. analyzed the thermodynamic performance of an AC system using a two-phase ejector as the expansion valve. Results show that the AC's coefficient of performance (COP) can be improved by over 15% [8]. Quansheng et al. proposed to use a heat exchanger as an energy storage device, which contains pressurized refrigerant. Experimental results indicate that an improvement of 2% in fuel efficiency can be achieved under the SC03 driving cycle [9]. Developing localized heating and cooling AC is able to improve the energy efficiency through avoiding to waste cooling power on the vacant seats. In [10], the authors investigated the thermal comfort in the vehicle compartment and quantified the energy saving of the localized AC system, with optimized front and ceiling vents up to 20.8% and 30.2%, respectively. Rugh et al. developed an advanced glazing method to reflect solar radiation and reduce the total thermal load [11], and achieved great progress in AC energy preservation.

Improving the AC system efficiency from the control aspect is also promising, and is usually called as energy management. Heuristic PID control has been widely used in current AC systems because of its easy implementation, simple structure and low cost [12]. However, PID control usually exhibits poor performance because it assumes the AC operating dynamics is linear. A thoroughly exploited AC system model is nonlinear and complex, then fuzzy logic controllers are preferred, with fuzzy control rules which can be expressed in heuristic terms [13]. Li et al. employed a novel neural network to aid the optimization of fuzzy rules. Results show that the thermal comfort level is able to be further increased with decrease of the energy consumption [14]. Nevertheless, fuzzy logic control excessively depends on personal experience and trivial tuning.

Due to the ability of disturbance integration, dynamic control and constraint handling, model predictive control (MPC) is attracting more attentions in the automotive industry and control system of HVAC [15-17].

MPC determines the system inputs via receding horizon optimal control, based on an open-loop model which is generally called the prediction model. The most important ability of MPC is to enforce pointwise constraints in each step, while providing the real-time optimal control designer by adjusting the weights in the objective function to minimize the cost [17-19]. Huang compared the performance of a simulated variable air volume AC system controlled by MPC and PI [20]. Results show that MPC controller demonstrates better transient response and is more robust than PI control. Moroşan studied the building zone temperature regulation with decentralized, centralized, and distributed MPC [21], demonstrating that decentralized MPC can reduce approximately 5.5% of the energy consumption, whereas centralized MPC and distributed MPC are able to achieve 36.7% increase in the thermal comfort and 13.4% reduction in the energy consumption. Afram

This work was supported by the National Natural Science Foundation of China (Grant No.51675042, 51705019) and China Postdoctoral Science Foundation funded project (2016M600049, 2017T100040).

The authors are with the National Engineering Laboratory for Electric Vehicles, School of Mechanical Engineering, Beijing Institute of Technology, Beijing 100081, China. *Corresponding author. (e-mail: hwhebit@bit.edu.cn; bit3jiahui@126.com; sunfch@bit.edu.cn; sunchao1988@163.com).

developed artificial neural network based the residential HVAC systems model and applied supervisory MPC to reduce operating cost of HVAC system [22]. The research indicates that MPC consumes more electricity but leads to lower operating cost because of its ability to keep the energy in building mass during off-peak hours [23]. Lefort proposed a hierarchical MPC (H-MPC) for energy consumption reduction in a residential house [24]. HMPC was composed of a scheduling MPC (S-MPC) and a piloting MPC (P-MPC). The S-MPC dealt with the slow- moving dynamics and the varying price of the electricity. The P-MPC managed the disturbances and fast-moving dynamics. Compared with a centralized MPC, the H-MPC demonstrated superior performance in terms of dissatisfaction cost. Pino et al. analyzed the behavior of AC system controlled by MPC in a fuel cell car under different driving cycles [25]. An increment of hydrogen consumption between 3% and 12.1% was found when the AC is turned on.

However, the AC system/plant's sensitivity to some of the key influence factors has not been fully studied, such as the solar radiation on the vehicle body, ambient temperature around the vehicle, and relative air flow speed. The above three factors are intermittent, uncertain and important disturbances to the AC plant, and have great influence on the controller performances. The main contributions made in this paper include:

- An AC plant model for EV sedan is established, and a comprehensive analysis of its electricity consumption sensitivity to solar radiation, external air temperature and relative air flow speed is presented;

- A stochastic model predictive controller (SMPC) is developed to realize efficient AC plant control, with the Markov-chain based velocity predictor adopted to provide velocity references in the control horizon;

- Real solar radiation, ambient temperature and air flow speed data are collected to validate the proposed SMPC, with comparison with dynamic programming and a bang-bang PID controller from the energy consumption and cabin temperature comfort perspectives.

The remainder of the paper is organized as follows: the AC system and thermal load model is established in Section II; Section III formulates the optimal AC power control problem with SMPC; DP and a rule-based bang-bang controller are described in Section IV; Section V presents the simulation results with conclusions given in Section VI.

## II. AIR CONDITIONING SYSTEM AND THERMAL LOAD MODEL

A scheme of the AC model established in this paper is demonstrated in Fig. 1. A model of the thermal load system takes the relative air flow velocity around the vehicle body, and the ambient conditions as inputs. The cabin temperature and the thermal loads ($Q$) to be balanced in the cabin are calculated through thermal superposition. In this paper, we assume the absolute air flow velocity outside of the vehicle is zero, so that the relative air flow velocity equals to the vehicle velocity.

The target of AC system controller is to maintain the thermal balance of the cabin air. Fig. 1(a) shows the main factors that influence the inside vehicle thermal condition, including solar flux, heat conduction, occupants in the car, and air ventilation. The thermal loads from the above sources are conducted into the vehicle cabin, causing temperature comfort affects. Fig. 1(c) illustrates that the air conditioner provides cool air to neutralize unexpected thermal loads, eventually to keep the cabin temperature within a comfortable range. The operation of the AC system is greatly affected by the ambient conditions. The cooling capacity ($Q_{\text{cool}}$) is a feedback response to the thermal load system (to be explained in Subsection II-A).

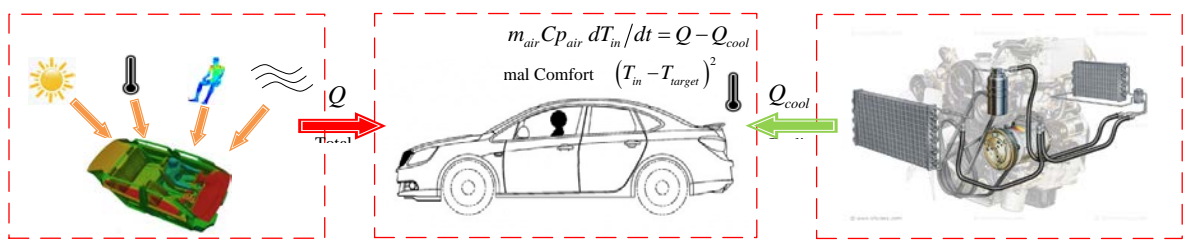

Fig. 1. Structure of thermal load model and air conditioning system

TABLE I
PARAMETERS OF THE AC SYSTEM

| Parameter | Value |
| --- | --- |
| Range of rotation speed | 1500-6500 r/min |
| Nominal capacity | 6.8 kW |
| Displacement | 36 cc/r |
| Air flow through evaporator | 0.186 kg/s |

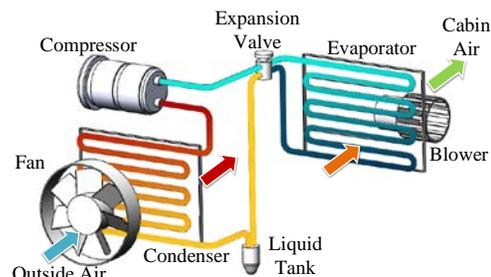

Fig. 2. Air conditioning system scheme.

## A. Air Conditioning Plant Model

An independent electrical air conditioner is selected in this paper. The parameters of the AC system are shown in Table I. The compressor drives the refrigerant (R-134a) to flow inside the AC system circularly to create a comfortable temperate in the cabin. The refrigerant flows through the evaporator to absorb heat and transfers it to the environment via condenser.

The AC plant scheme is illustrated in Fig. 2, where the compressor is the main energy-consuming unit. The coefficient of performance (COP) of the air conditioner is defined as the ratio between the cooling capacity and the power consumed by the AC system. COP is influenced by cabin temperature, ambient temperature and partial load ratio (PLR), as shown in Fig. 3 (a-b), respectively [25]. PLR is the ratio between the actual cooling capacity and the nominal capacity of AC system in operating conditions. The COP generally increases as the cabin temperature increases or the ambient temperature decreases. When the PLR is between 0.4 and 0.8, the COP varies only slightly. The power consumption required from AC system is formulated as,

$$P = \frac{Q_{\text{cool}}}{COP} \quad (1)$$

where $P$ is the power consumption, and $Q_{\text{cool}}$ is the cooling capacity of the AC system.

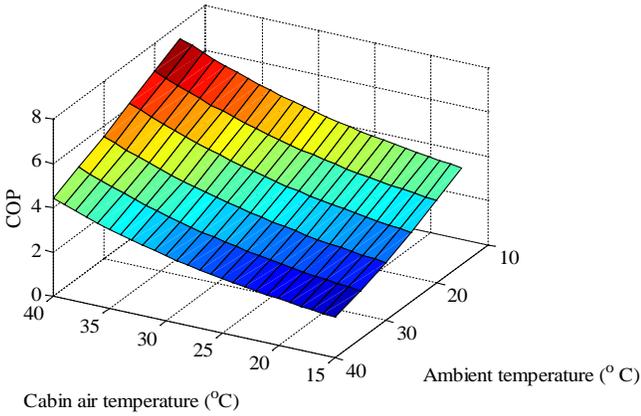

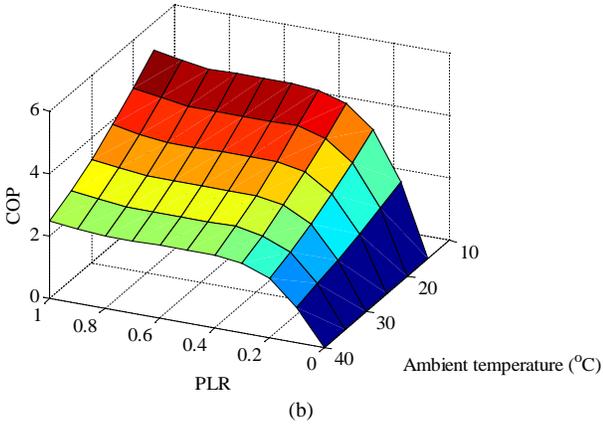

Fig. 3. The COP used in our study. (a) COP vs the cabin air temperature and environment temperature; (b) COP vs the partial load ratio.

## B. Thermal Load Model

External thermal loads transfer heat into the vehicle cabin mainly through two ways: conduction and radiation. The thermal conduction load is caused by the difference of the temperature between the cabin and the ambient environment. The thermal radiation load is caused by the incident solar radiation. Internal thermal loads include: the heat supplied by the car occupants, and the heat brought from the ventilation system.

1) Conduction thermal load $Q_c$: the conduction thermal load contains the heat which transfers into the cabin through the vehicle body and the windows. The opaque vehicle body, including roof panel, front wall, rear wall, floor and side-parts, is approximated to a multi-layer structure model in order to simplify the system. The thermal conduction load can be calculated as follows [23]:

$$Q_c = \sum KF\left[\left(T_{\text{out}} + \frac{\rho I}{\alpha_w}\right) - T_{\text{in}}\right] \quad (2)$$

where $Q_c$ is the thermal load caused by conduction, $K$ is the heat transfer coefficient, $F$ is heat transfer area of the corresponding envelope, $T_{\text{out}}$ is the ambient temperature, $\rho$ is the mean thermal absorptivity of the surface, $I$ is the density of incident solar flux, $\alpha_w$ is the convective heat transfer coefficient, $T_{\text{in}}$ is the cabin air temperature.

Importantly, the heat transfer coefficient $K$ varies from different materials. In our paper, according to heat transfer theory, $K$ in equation (2) is calculated by:

$$K = \left(\sum \frac{\delta_i}{\lambda_i} + \frac{1}{\alpha_w} + \frac{1}{\alpha_n}\right)^{-1} \quad (3)$$

for all i, where $\delta$ is the thickness of the material, $\lambda$ is the thermal conductivity of the material, $\alpha_n$ is the convective heat transfer coefficient. $\alpha_w$ is related to the vehicle velocity, air flow velocity and flow direction. The flow field of the body is instable in a wide range of vehicle speeds. In this paper, $\alpha_w$ and $\alpha_n$ are generally calculated by empirical equation (4) and (5), respectively [15]:

$$\alpha_w = 1.163(4 + 12\sqrt{v}) \quad (4)$$

$$\alpha_n = \begin{cases} \begin{cases} a + b \cdot \Delta t_b, \Delta t_b < 5°C \\ c \cdot \Delta t_b^{0.25}, \Delta t_b > 5°C \end{cases} & v_{\text{air}} \in [0.25, 0.5]\, m/s \\ 8.7 \sim 29, & v_{\text{air}} \in [0.5, 3]\, m/s \end{cases} \quad (5)$$

where $v$ is the wind speed relative to the vehicle, which we assume equals to the vehicle velocity, $a = 3.49$, $b = 0.093$, $c$ values 2.67~3.26 are empirical coefficients, $\Delta t_b$ is the temperature difference of internal vehicle surfaces and cabin air, $v_{\text{air}}$ is the speed of cabin air.

2) Radiation thermal load $Q_r$: the radiation thermal load is mainly caused by solar flux. We assume that the horizontal component of solar radiation is small enough to be neglected. Thus, the radiation load $Q_r$ can be calculated as follows [26]:

$$Q_r = \sum_{i=1}^{n} q_i F_{g,i} \sin \varphi_i \quad (6)$$

$$q_i = \left(\eta I + \rho_g I \frac{\alpha_n}{\alpha_w}\right) C \quad (7)$$

where $q$ is thermal load of window caused by solar radiation in the horizontal direction per $m^2$, $F_g$ is the area of the window, $\varphi$ is the angle between the window and the vertical direction, $\eta$ is the input coefficient of solar radiation through the window, $\rho_g$ is the mean absorptivity of the window and $C$ is shading correction factor.

3) Sensible heat supplied by the car occupants $Q_m$: it is affected by gender, age, labor intensity and so on. Based on the experience, the sensible heat is about 145 W supplied by the driver and 116 W supplied by each passenger [27]. Considering the effects of proportion of occupants, a correction factor is used to estimate the sensible heat [26], see equation (8):

$$Q_m = 116 n_p \beta + 145 \qquad (8)$$

where $n_p$ is the number of passengers and equals 4 in this paper, $\beta$ is the correction factor.

4) Heat brought from the ventilation system $Q_n$: parts of thermal loads are generated resulting from the ventilation of the AC system. The ventilation thermal load $Q_n$ is given by

$$Q_n = m_e(1-\xi)Cp_{air}(T_{out} - T_{in}) \qquad (9)$$

where $m_e$ is defined as the mass of air flow through the evaporator and equals $0.186\ kg/s$, $\xi$ is air recirculation coefficient (not all the air introduced in the cabin is from outside) and $Cp_{air}$ is heat capacity of indoor air [26].

Overall, the temperature of cabin air can be calculated as follows:

$$\rho_{air} V_{air} C p_{air} \frac{dT_{in}}{dt} = Q_c + Q_r + Q_m + Q_n - Q_{cool} \qquad (10)$$

where $\rho_{air}$ and $V_{air}$ are density and volume of the cabin air respectively.

### III. STOCHASTIC MODEL PREDICTIVE CONTROLLER

#### A. Model Predictive Control

Model predictive control is a popular strategy which has been widely employed in industry as an effective method of solving constrained nonlinear control problems [28, 29]. The purpose of MPC is to obtain the control sequences by solving an open-loop finite horizon optimal control problem at each sample time according to a prediction model of the process by regarding the current state as the initial prediction state. The control inputs are implemented in accordance with a receding horizon scheme. The model uncertainties and disturbances, which are often entirely ignored in the prediction process, actually highly influence the overall performance. Robust MPC schemes that deals with the disturbances are mostly according to the min-max approach, where the performance index to be minimized is computed over the worst possible disturbance realization [30]. However, nominal controllers which ignore the disturbances may result in poor performances when used in practical processes.

In this paper, we adopt stochastic model predictive control for the AC system power control. The optimal energy management problem is solved via DP at each time step, and the statistical information of the disturbances is exploited to minimize the performance index [31, 32]. In the SMPC formulation, the cabin air temperature is selected as the first state variable. In order to avoid the compressor varying operating conditions too frequently, the cooling capacity of AC system is selected as the second state variable and also the control variable. Denoting $x$ as the state variable, $u$ as the control variable and $y$ as the output, the proposed control-oriented AC system model can be formulated as

$$\begin{cases} \dot{x}_1 = \dfrac{Q_c + Q_r + Q_m + Q_n - u}{\rho_{air} V_{air} C p_{air}} \\ \dot{x}_2 = \dot{u} \\ y = x_1 \end{cases} \qquad (11)$$

with $x_1 = T_{in}$, $x_2 = Q_{cool}$, $u = Q_{cool}$.

For all simulations, time step is $\Delta t$. At time $k$, the cost function $J_k$ is formulated as,

$$J_k = \int_{k\Delta t}^{(k+H_p)\Delta t} \left(\omega_1 P_{AC} + \omega_2 (T_{in} - T_{target})^2\right) dt \qquad (12)$$

where $H_p$ is the prediction horizon length and equals to the control horizon length, $\omega_1$ and $\omega_2$ are the weight coefficients that determine the importance of electric power and the temperature error, respectively. The temperature difference between the cabin temperature $T_{in}$ and the target temperature

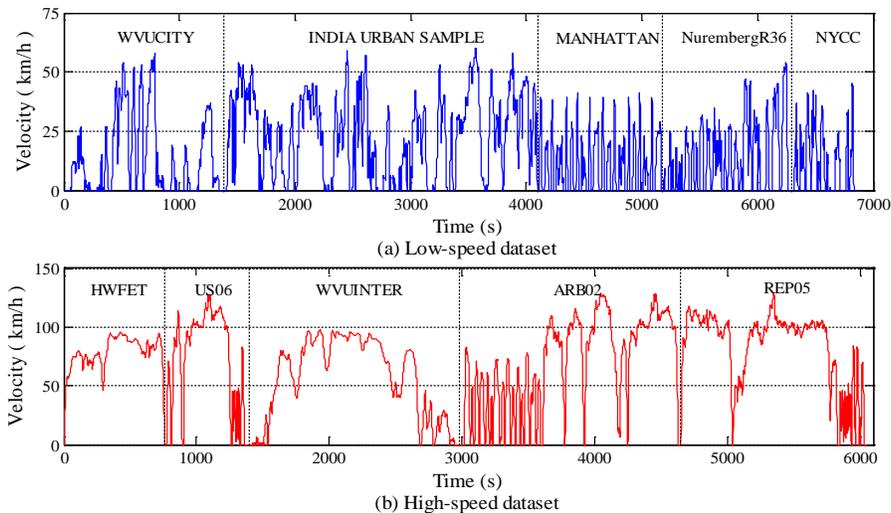

Fig. 4. Speed sample dataset used in prediction model. (a) Low-speed dataset; (b) High-speed dataset.

$T_{\text{target}}$ is used to measure the cabin comfort.

At the same time, the following physical constraints must be respected:

$$\begin{cases} \omega_{c_{\min}} \leq \omega_c \leq \omega_{c_{\max}} \\ Q_{\text{cool}_{\min}} \leq Q_{\text{cool}} \leq Q_{\text{cool}_{\max}} \\ |\dot{Q}_{\text{cool}}| \leq 500 \end{cases} \quad (13)$$

SMPC is used in the supervisory level of the control structure. At each time step, the optimization problem is solved by DP with the vehicle speed sequence predicted by Markov chain [33]. The optimization control procedure can be described as:

(1) Regarding the current vehicle speed $v_k$ as the initial prediction state, the Markov chain exploits the statistical information of the historical speed to predict the future speed $v_{k+1}$. Then $v_{k+1}$ is considered as the new initial state to predict $v_{k+2}$. In this way, the predictive vehicle velocity sequence $V_{\text{pre}} = \{v_{k+1}, v_{k+2}, \cdots, v_{k+H_p}\}$ can be generated.

(2) SMPC controller calculates the optimal control decisions minimizing the cost function (12).

(3) Implement the first item of the optimal control signals, feedback the system states, and repeat the control procedure.

*B. Stochastic Prediction of the Vehicle Speed*

It should be noted that the future vehicle velocity during the control horizon is predicted by Markov-chain at each time instant [34]. The solar radiation and environment temperature can be obtained from the online weather service. These three kinds of information during each control horizon are important demand inputs to the SMPC controller, with the optimization problem solved in real time at each time step [35].

In this paper, the vehicle velocity is modeled as a Markov chain which is defined by an emission probability matrix $T_M \in \mathbb{R}^{p \times q}$ such that

$$T_{ij} = P[V_{k+1} = v_j | V_k = v_i] \quad (14)$$

where $T_{ij}$ is the $(i, j) - $ th item of the emission probability matrix, $P[x]$ denotes the probability of $x$.

Emission probabilities $T_{ij}$ are generated from the sample dataset. Once the driving cycle sample is known, the emission probabilities will be determined at the same time. Generally, the sample dataset consists of standard driving cycles or the past velocity records. In order to compare the performance of SMPC with DP and bang-bang controller, the model is simulated under the low speed and high-speed conditions respectively. As shown in Fig. 4, the low speed sample dataset is composed by five standard driving cycles (WVUCITY, INDIA URBAN, MANHATTAN, NurembergR36 and NYCC). A different set of driving cycles (HWFET, US06, WVUINTER, ARB02 and REP05) compose the high-speed sample dataset.

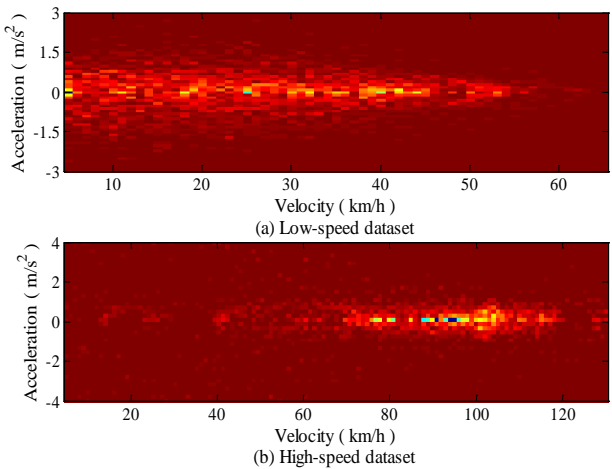

Fig. 5. Velocity-acceleration diagrams of the low- and high-speed datasets. The low speed is shown at the top. The red-scale shows the density function of velocity and acceleration.

A velocity-acceleration diagram for low and high-speed data is shown in Fig. 5. $T_{ij}$ can be calculated based on the statistical frequency information of the sample dataset:

$$T_{ij} = \frac{n_{ij}}{n_i} \quad (15)$$

where $n_{ij}$ is the number of times that the emission from $v_i$ to $v_j$ appears, $n_i$ is the total number of appearances times of $v_i$. The emission probability matrix extracted from low speed driving cycle sample is shown in Fig. 6 with $p = 60$ and $q = 45$.

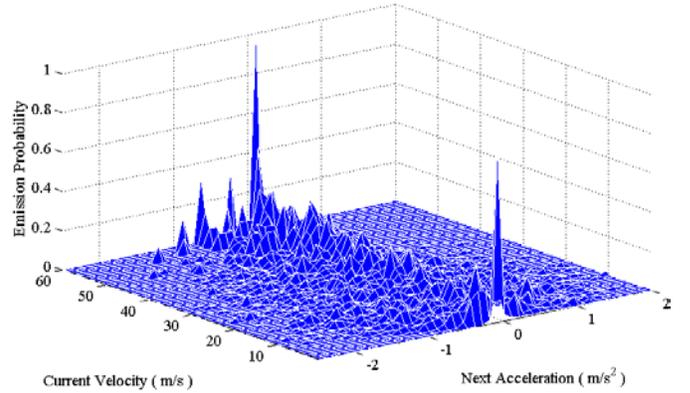

Fig. 6. Markov emission probability matrix of the low speed sample dataset.

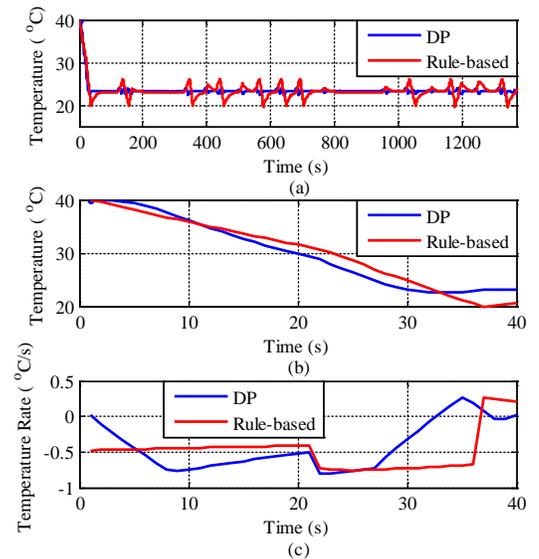

Fig. 7. (a) Cabin temperature control results calculated via DP and rule based strategy; (b) cabin temperature in the first 40 seconds; (c) cabin temperature rate in the first 40 seconds.

## IV. Dynamic Programming and Rule-based Control

In order to evaluate the performance of SMPC, we use DP and a rule-based bang-bang controller to compare with SMPC under the same operating conditions.

DP is often used to solve nonlinear optimization problems [35], and serves as benchmarks. Minimizing the total AC system electricity consumption via DP is formulated as,

$$J_N = \int_0^{N\Delta t} \left( \omega_1 P_{AC} + \omega_2 (T_{in} - T_{target})^2 \right) dt \tag{16}$$

$$\dot{x}_1 = f(x, u, d) \tag{17}$$
$$\dot{x}_2 = h(x, u, d) \tag{18}$$

where $f(x, u, d)$ and $h(x, u, d)$ are the state dynamics defined in equation (11).

The rule-based bang-bang controller can also maintain the cabin temperature within a comfortable range. According to the operating rule, the compressor mainly works as the following:

$$Q_{cool} = Q_{cool\_max}, (T_{in} \geq T_{target\_high}) \tag{19}$$

$$Q_{cool} = k_{rule} \frac{T_{in} - T_{target_{low}}}{T_{high} - T_{target_{low}}} + b_{rule}, (T_{target\_low} \leq T_{in} \leq T_{target\_high}) \tag{20}$$

$$Q_{cool} = Q_{cool\_min}, (T_{in} \leq T_{target\_low}) \tag{21}$$

where $T_{target\_high}$ and $T_{target\_low}$ are the upper and lower limits of the target temperature range, $k_{rule}$ (=1000) and $b_{rule}$ (=2000) are the proportion coefficient and the interception, respectively. They are determined via trial and errors.

When the cabin temperature is higher than the upper limit of the target temperature, the compressor runs under the largest cooling capacity to reduce the vehicle cabin temperature to the lower limit as fast as possible; when the temperature drops below the lower limit of the target temperature, the cooling capacity is controlled along function as shown in equation (20). Therefore, the cabin temperature will be maintained within the target temperature range reasonably.

To test the performance of DP and the bang-bang controller, the AC system were simulated with the two controllers under UDDS driving cycle. The solar radiation is 1200 W and the ambient temperature is 35°C. The results are shown in Fig. 7. The DP target temperature is set as 23 °C and the target temperature range of the bang-bang controller is set as 20~26°C. Both of the two controllers can maintain the cabin temperature within the target range. As shown in Fig. 7 (b) and (c), the temperature changing rate of DP is greater than the rule-based controller in the first 20 seconds, but smaller during time 27s to 37s. The energy consumption will be systematically compared in the next section.

## V. Simulation Results

### A. Sensitivity Analyses on AC Performance via DP

According to thermal load model, we can find that the total thermal load is mainly affected by the relative air flow velocity, solar radiation and ambient temperature. In the following subsections, a detailed analysis of the influence caused by these three factors is performed based on DP.

#### 1) Influence by Relative Air Flow Velocity

Relative air flow speed (vehicle velocity) directly impacts the conduction thermal load $Q_c$ and the radiation thermal load $Q_r$. A lower relative air speed leads to a heavier thermal load. In order to study the relationship between the electricity consumption of AC system and the relative air flow speed, we simulated the AC model at different speeds with the solar radiation of 900W and the ambient temperature of 30°C. In all simulations of Section V-A, the cabin air temperature was maintained a constant value $T_{target}$ (20°C, 22°C, 24°C or 26°C) from beginning to end. The results are shown in Fig. 8.

It can be seen that the electricity consumption of AC system changes drastically in low-speed range and is basically flat in high-speed range. In addition, the results show that there is a linear relationship between the AC consumption and the target temperature under the same driving velocity. The consumption drops by approximately 20% from $v = 0$ to $= 5\ km/h$. As the velocity increases, the consumption trajectory becomes increasingly gentler. The AC system that operates under 100 $km/h$ consumes only 1~2% less than 40 $km/h$. Therefore, we can consider that when the relative speed is lower than 5 $km/h$, the influence on consumption is dramatic; when the speed is between 5 $km/h$ and 40 $km/h$, the consumption decreases with increasing speed and the decrement can't be neglected; when the speed is higher than 40 $km/h$, the consumption is essentially unchanged. In other words, the traffic jams or a windless weather can result in AC system having a large proportion in consume of electricity.

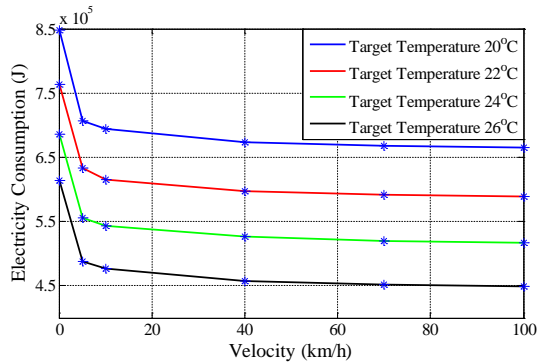

Fig. 8. Electricity consumption of AC system at different air relative velocities and cabin air temperatures.

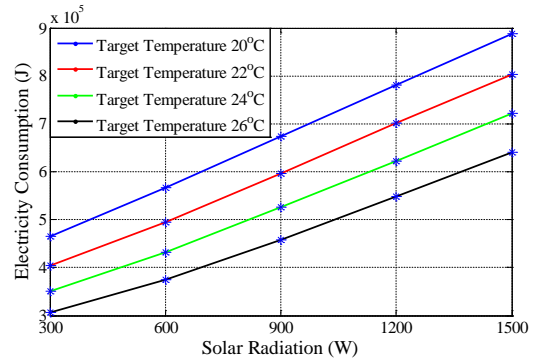

Fig. 9. Electricity consumption of AC system under different solar radiations and cabin air temperatures.

*2) Influence by Solar Radiation*

Besides the relative air flow speed, the conduction thermal load $Q_c$ and the radiation thermal load $Q_r$ are also related to the solar radiation. The stronger solar radiation is, the heavier thermal load is generated. Therefore, the model was simulated under different solar radiation conditions with the relative air speed of 40 $km/h$ and the ambient temperature of 30℃. The simulation results are shown in Fig. 9.

As can be seen from Fig. 9, the results show that AC consumption has a nearly linear relationship with the solar radiation, because radiation thermal load and a part of conduction thermal load are linearly associated with solar radiation. These two items compose the main component part of the total load. During summer season, the radiation is often between 700-1300 W. One degree reduction of the temperature setting will increase AC electricity consumption by nearly 7%. That is, we can conserve energy at expense of the comfort in some harsh conditions. In addition, in order to reduce the radiation thermal load $Q_r$, we can use the solar reflecting glass windows which can reflect 83% of the infrared solar and allow only 3% of transmission of the solar flux into the cabin. Such high solar reflectivity of the solar reflecting glass can result in a much lower cabin air temperature and energy consumption.

*3) Influence by Ambient Temperature*

Ambient temperature greatly influences the conduction thermal load $Q_c$, the ventilation thermal load $Q_n$, and also the AC operating conditions. Air with higher ambient temperature can transfer heat into the cabin much more quickly. In addition, the COP of AC system decreases with the increase of ambient temperature. That means the AC system consumes more electricity to generate the same cooling capacity with a small COP. It is necessary to conduct the simulations under different ambient temperatures.

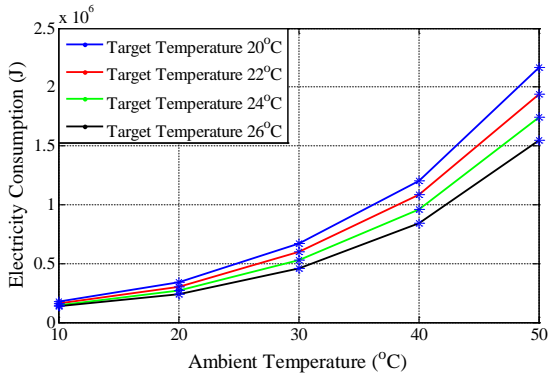

Fig. 10. Electricity consumption of AC system under different ambient temperatures and target cabin air temperatures.

In all cases, the relative air speed is 40 $km/h$ and the solar radiation is 900 W. The simulation results are shown in Fig. 10. We can see that the AC consumption increases exponentially with ambient temperature. The consumption caused by the target temperature grows with the ambient temperature. In high temperature conditions, the influence of ambient temperature becomes considerably pronounced. The application of the body thermal insulation is an effective measure to reduce conduction load when the cabin temperature is within a comfortable range. But it also leads to a much higher initial cabin temperature when the vehicle soaks in the sun.

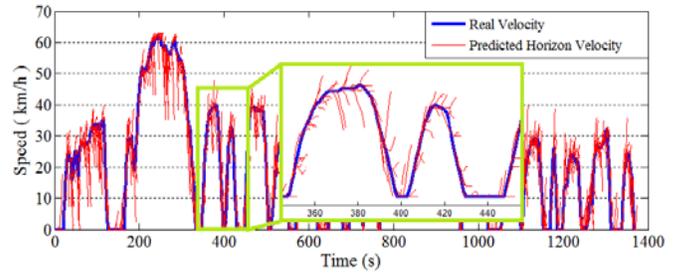

Fig. 11. Velocity prediction results of Markov chain prediction.

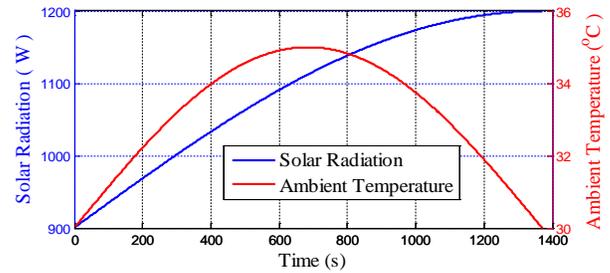

Fig. 12. Solar radiation and ambient temperature trajectories.

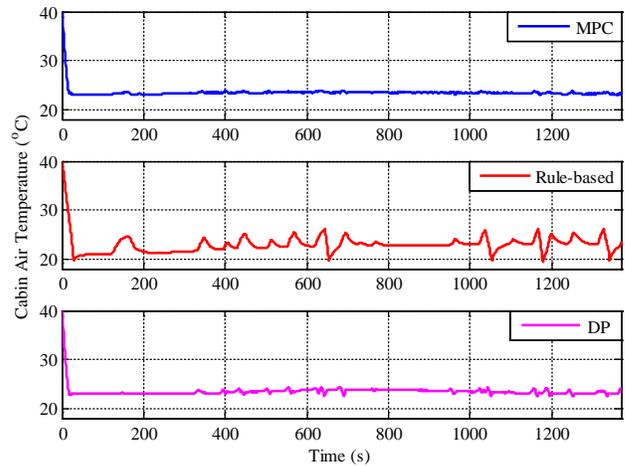

Fig. 13. Cabin air temperature trajectories of SMPC, rule-based and DP controllers under low-speed test cycle.

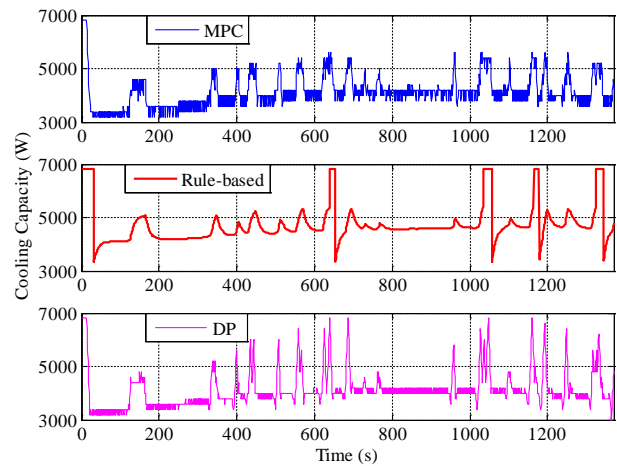

Fig. 14. Cooling capacity trajectories of SMPC, rule-based and DP controllers under low-speed driving cycle.

TABLE II
SIMULATION RESULTS OF SMPC, BANG-BANG AND DP CONTROLLERS UNDER STANDARD DRIVING CYCLES

| Controller | Cycle | Energy× $10^6$ (J) | Saving | Mean Temperature (℃) | Standard Deviation (℃) |
| --- | --- | --- | --- | --- | --- |
| SMPC | Low-speed | 2.0242 | 14.58% | 23.19 | 0.97 |
|  | High-speed | 1.9950 | 15.27% | 23.18 | 0.98 |
| Bang-bang | Low-speed | 2.3698 | — | 23.36 | 1.76 |
|  | High-speed | 2.3545 | — | 23.22 | 1.76 |
| DP | Low-speed | 2.0157 | 14.94% | 23.18 | 1.02 |
|  | High-speed | 1.9899 | 15.49% | 23.17 | 1.02 |

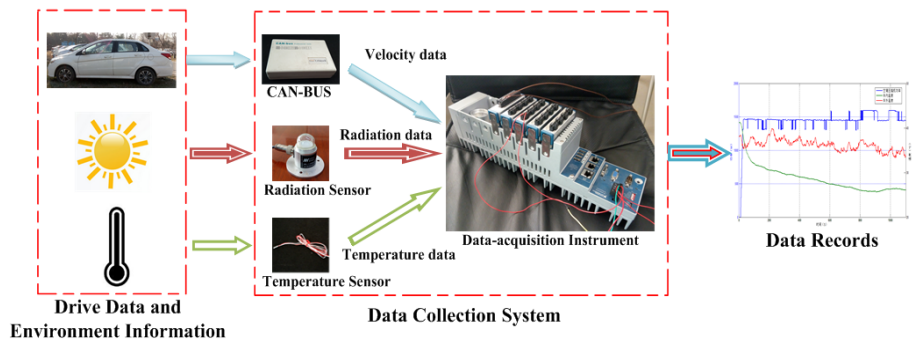

Fig. 15. Driving and environment data measurement setup.

### B. AC Controller Comparison – Standard Driving Cycle

This section provides a comprehensive comparison of SMPC, bang-bang and the DP controllers. The performance of the three controllers is compared in electricity consumption and cabin air temperature. Two driving cycles are used for the simulation: UDDS×0.68 and UDDS×1.45, which represent the low-speed and high-speed driving cycles respectively.

The control horizon length of SMPC is set as 5 seconds. With the sample dataset mentioned in Section III-B, taking the low-speed driving cycle as an example, the velocity prediction results of Markov chain can be seen visually in Fig. 11. In general, the predictive velocities can follow the actual velocity trajectory very well in the first several seconds of the prediction sequence. The prediction length is the same as the SMPC control horizon length. Fig. 12 is the assumed solar radiation and ambient temperature.

Fig. 13 plots the cabin air temperature trajectories simulated using SMPC, rule-based and DP, respectively. The initial cabin temperature is set as 40℃. The target temperature of SMPC and DP is 23℃, and the target temperature range of rule-based controller is 20~26℃. We can see all the controllers can result in a rapid drop in temperature at first in order to create a comfortable environment as soon as possible. Then SMPC and DP can maintain the cabin air temperature around the target temperature successfully. The temperature with rule-based controller oscillates between 20℃ and 26℃ periodically based on the bang-bang control law. Unexpectedly, the temperature fluctuation of SMPC is even less than DP. The reason could be that DP is a global optimization; whereas the MPC controller yields a locally optimal solution for each control horizon, and lacks flexibility.

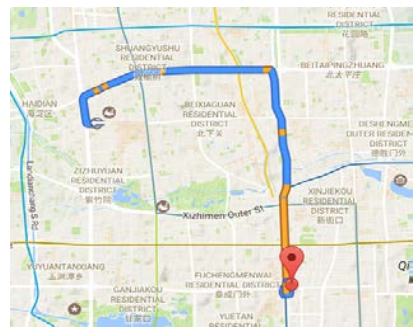

Fig. 16. Real driving route for data collection in this paper.

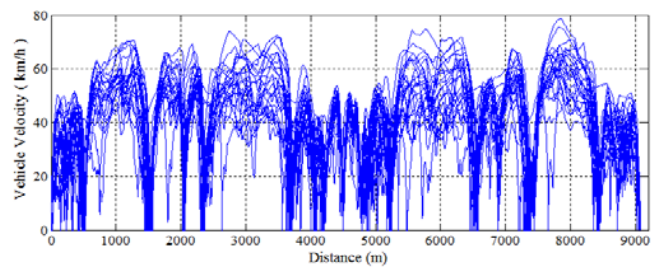

Fig. 17. The 25 measured velocity trajectories.

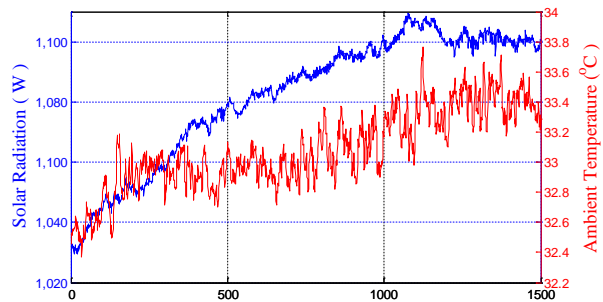

Fig. 18. The measured solar radiation and ambient temperature trajectories.

TABLE III
SIMULATION RESULTS OF SMPC, RULE-BASED AND DP CONTROLLERS UNDER TEN TEST CYCLES

| Cycles | Controller | Energy× $10^6$ (J) | Saving (%) | Mean Temperature (°C) | Standard Deviation (°C) |
|---|---|---|---|---|---|
| UDDS | SMPC | 2.00 | 14.16 | 23.21 | 0.99 |
|  | Rule-based | 2.33 | — | 23.38 | 1.86 |
|  | DP | 1.97 | 15.39 | 23.46 | 0.98 |
| CSHVR | SMPC | 2.65 | 14.11 | 23.52 | 0.83 |
|  | Rule-based | 3.09 | — | 23.74 | 1.68 |
|  | DP | 2.59 | 15.94 | 23.44 | 0.86 |
| LA92 | SMPC | 2.10 | 13.68 | 23.23 | 0.96 |
|  | Rule-based | 2.43 | — | 23.57 | 1.72 |
|  | DP | 2.07 | 15.05 | 23.48 | 0.95 |
| NurembergR36 | SMPC | 1.66 | 13.35 | 23.28 | 1.01 |
|  | Rule-based | 1.91 | — | 23.33 | 2.10 |
|  | DP | 1.64 | 14.41 | 23.50 | 1.01 |
| India HWY | SMPC | 1.23 | 12.02 | 23.17 | 1.13 |
|  | Rule-based | 1.40 | — | 23.64 | 2.37 |
|  | DP | 1.21 | 13.33 | 23.39 | 1.12 |
| 1st Test Cycle | SMPC | 1.64 | 11.54 | 23.17 | 1.00 |
|  | Rule-based | 1.86 | — | 23.38 | 1.89 |
|  | DP | 1.60 | 13.64 | 23.02 | 1.14 |
| 2nd Test Cycle | SMPC | 1.70 | 11.59 | 23.18 | 0.95 |
|  | Rule-based | 1.93 | — | 23.30 | 1.82 |
|  | DP | 1.66 | 13.74 | 23.03 | 1.09 |
| 3rd Test Cycle | SMPC | 1.64 | 11.63 | 23.17 | 0.94 |
|  | Rule-based | 1.86 | — | 23.31 | 1.92 |
|  | DP | 1.61 | 13.69 | 23.03 | 1.15 |
| 4th Test Cycle | SMPC | 1.68 | 11.65 | 23.17 | 1.06 |
|  | Rule-based | 1.91 | — | 23.30 | 1.94 |
|  | DP | 1.65 | 13.52 | 23.03 | 1.13 |
| 5th Test Cycle | SMPC | 1.89 | 11.50 | 23.18 | 0.89 |
|  | Rule-based | 2.13 | — | 23.29 | 1.80 |
|  | DP | 1.84 | 13.58 | 23.05 | 1.06 |

Table II summarizes the simulation results, where can be seen that the electricity consumption of AC system and the mean cabin air temperature show similar result patterns among the three controllers. The performance of SMPC is very close to DP. The mean cabin temperatures of SMPC and DP equal to 23.2°C in both rest cycles, but the standard deviation of DP is slightly larger than that of SMPC. SMPC consumes 14.58% and 15.27% less electricity than the rule-based controller in the two tests, respectively. The rule-based controller produces the highest mean temperature and also consumes the most power.

The cooling capacity trajectories of the three controllers are shown in Fig. 14. This figure elucidates how SMPC and DP consume less electricity and keep a more stable cabin air temperature than the rule-based controller. At the beginning of the cycle, all the three controllers operate AC system producing a high cooling capacity. Then the cooling capacities of SMPC and DP decrease fast as the cabin air temperature until it is close to the target temperature and keep a stable value which changes with the varying environmental conditions with a very small fluctuation.

However, the cooling capacity of the rule-based controller varies greatly as the changing cabin temperature. When the vehicle speed is lower than a certain value which is also related to the environment, the cooling capacity of SMPC increases rapidly and stabilizes around a higher value. Unlike SMPC, DP algorithm operates the compressor in a wider range of cooling capacity with small comfort cost in order to minimize the energy consumption. In summary, the performance of SMPC is quite close to DP and seeks more reasonable operation behaviors than rule-based bang-bang control.

*C. AC Controller Comparison – Real Driving Cycle*

In order to test the performance of the three controllers, an experiment was performed to collect the driving data, solar radiation and ambient temperature in Beijing during three weeks in August 2016. The data was collected between 11:00 AM and 13:00 PM. Measurements were conducted on an electric vehicle with a data collection system as shown in Fig.15. The vehicle speed was read from the CAN-bus and recorded in the data acquisition system. The solar radiation and ambient temperature were collected with a radiation sensor and a temperature sensor which were placed on the roof of the car. In the future potential applications, the solar, wind and temperature conditions can be obtained from the Cloud.

The driver drove smoothly and the weather was sunny. The data was collected on the same route (as shown in Fig. 16) by repeating 25 times. Therefore, it is concluded that the data was collected under similar conditions. The measured velocity trajectories are shown in Fig. 17, and the real collected solar radiation and ambient temperature data are illustrated in Fig. 18. Arbitrarily selected five real driving cycles from Fig. 17, and another five standard driving cycles (UDDS, CSHVR, LA92, NurembergR36, India HWY) are used for the simulation. The rest 20 cycles are used to compute the Markov-chain emission probability matrix. The results are shown in Table III, where we can summarize that the proposed SMPC controller is able to maintain the vehicle cabin temperature at a comfortable level with much better electricity economy achieved.

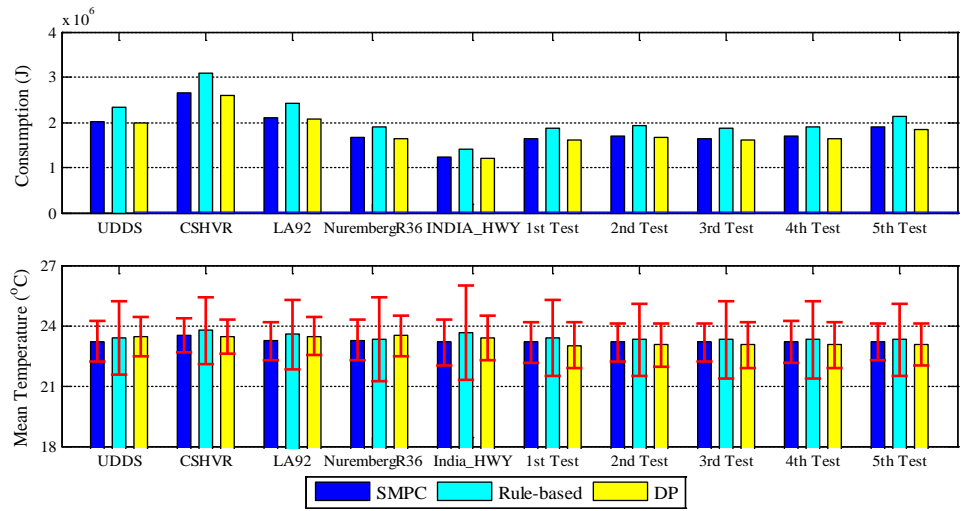

Fig. 19. Comparisons of AC consumption and mean cabin temperature with SMPC, rule-based and DP controllers.

Fig. 19 shows the comparisons of electricity consumption and mean cabin air temperature in all cases. DP maintains the temperature around the target temperature and consumes the least electricity. SMPC produces a comfortable cabin temperature which is very close to DP. However, the energy consumption is 1.24%~2.50% more than that of DP.

Rule-based bang-bang controller performs the worst both in the comfort and energy performances. In most cases, the cabin air temperature of bang-bang control is higher than the other two controllers and fluctuates greatly. The energy consumption is always 15.38%~18.97% more than SMPC or DP.

## VI. CONCLUSIONS

A sensitivity study of the energy consumed by the air conditioning system in EVs under different solar radiation, ambient temperature and vehicle velocity conditions is conducted in this paper. A comprehensive analysis based on DP simulation indicates that the electricity consumed by the AC system has a nearly linear relationship with the solar radiation, but increases exponentially with the ambient temperature. Outside air flow speed also influences the inside thermal load balance greatly, especially when the vehicle velocity is below 40km/h. As the vehicle speeds up, the influence fades.

Stochastic model predictive control is proposed in this paper to enhance the energy efficiency of AC systems, with Markov-chain developed to predict the future vehicle velocity references during each control horizon. A rule-based bang-bang controller and dynamic programming offline control method are adopted to evaluate the SMPC. Experiments are conducted to collect real solar, temperature and velocity data. Comparison results demonstrate that SMPC controller performs very close to DP on the total energy consumption and cabin comfort. The energy economy is improved by 11.50%~14.16% than the rule-based controller, and the cabin temperature fluctuation is reduced by over 50.4%, resulting with a much better cabin comfort.

This study provides better understanding of the energy expense caused by AC systems in electric vehicles, and the proposed SMPC controller is able to improve the EV driving range by about 3.6%. Future work includes further application of the control method to real air-conditioning plants.